\documentstyle[twocolumn,prb,aps,epsfig]{revtex}

\begin{document}
\title{ On the nature of the spin-singlet ground state in CaCuGe$_2$O$_6$ 
         }

\author{Roser Valent\'\i$^1$, T. Saha-Dasgupta$^2$ and C. Gros$^1$ } 

\address{$^1$Fakult\"at 7, Theoretische Physik,
 University of the Saarland,
66041 Saarbr\"ucken, Germany.}

\address{$^2$S.N. Bose National Centre for Basic Sciences, 
JD Block, Sector 3,
 Salt Lake City, Kolkata 700098, India.}

\date{\today}
\maketitle

\begin{abstract}

 We investigate by means of {\it ab initio} electronic structure
analysis and Quantum Monte Carlo calculations
the scenario where longer-ranged magnetic
interactions dominate over short-ranged interactions in
the physical description of compounds.  This question is 
discussed, in particular, for the case of  CaCuGe$_2$O$_6$ which 
shows a spin-singlet behavior induced by third nearest neighbor
copper pairs.

\end{abstract}


\vspace{1cm}

  There are a few low-dimensional materials which show, contrary
to initial intuitive arguments,  longer-ranged
dominated exchange couplings.   Examples include
systems like the most discussed 1/5-depleted
Heisenberg system  CaV$_4$O$_9$\cite{Taniguchi_95} or the alternating
 chain compound
(VO)$_2$P$_2$O$_7$\cite{Johnston_87}.   
CaV$_4$O$_9$ which was originally viewed as an array 
of weakly coupled square plaquettes of V$^{4+}$ ions,
each in a singlet in the ground state, is found to be a system 
with strongest coupling between next nearest neighbor
 vanadiums\cite{explanation_1} 
 as revealed by a recent electronic
 structure study\cite{Hellberg_99}.
 In
(VO)$_2$P$_2$O$_7$  the strongest exchange path\cite{Tennant_97} is found
 to be that between two
V$^{4+}$ ions  through two phosphate groups PO$_4$ and not 
 between 
 nearest neighbor vanadium ions V-O-V
as was initially thought\cite{Johnston_87}.
 From these examples, we learn that the subtle competition between the
various interactions can lead to surprises and a microscopic study
is needed.
 The detailed investigation of these systems often unveils interesting
properties.
One such a system is CaCuGe$_2$O$_6$, a system 
 related to the 
 spin-Peierls compound  CuGeO$_3$\cite{Hase_93}.

Susceptibility measurements\cite{sas95} on CaCuGe$_2$O$_6$ show the existence
 of a spin-singlet ground state with an energy gap of 6 meV.  But,
 contrary to CuGeO$_3$ where the spin-gap opens at the spin-Peierls
phase transition\cite{Hase_93},
 in CaCuGe$_2$O$_6$ the spin gap is intrinsic and 
there is no phase transition between 4.2K and 300K.  Also,
inelastic neutron scattering (INS) measurements were  carried out
 on CaCuGe$_2$O$_6$ powder and the existence of a finite spin-gap
 was confirmed\cite{Zheludev_96}.

\begin{figure}[t]

\vspace*{-0.2cm}

\centerline{\hspace*{0.5cm}
\epsfig{file=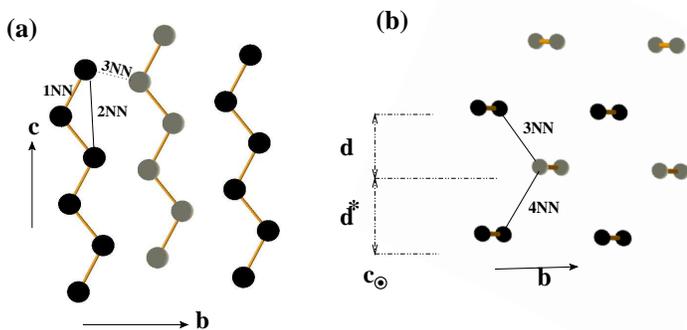,width=0.55\textwidth}
           }
\vspace*{0.0cm}
\caption{\label{structure1} 
(a) Projection on the $bc$ plane of the Cu zigzag
 chains  in   
CaCuGe$_2$O$_6$, where
 the 1NN, 2NN and 3NN Cu pairs are indicated. The chains alternate
 between two neighboring $bc$ plane (denoted by 
 black and grey balls).
 (b) Projection on the $ab$
 plane of the Cu sites, where the 3NN and 4NN Cu pairs are indicated. The
 linked Cu sites correspond to 1NN. $d$ and $d^*$ denote the distances between
two neighboring $bc$ planes. }

\end{figure}

  A question which neither INS nor susceptibility measurements could
  clarify was which Cu-pairs form the singlets in this
 material. This issue is of great importance since it determines
 the magnetic properties of the system.
 The structure of this material shows an obvious 
 zig-zag 1D chain of spin=1/2 Cu$^{2+}$ ions along
 the c-axis (see Fig.\ \ref{structure1}).
  Nevertheless, the magnetization and susceptibility data
of Sasago {\it et al.}\cite{sas95}
 are in disagreement with a spin=1/2 Heisenberg chain model.
These authors proposed that the spin-gap behavior of the system should
be due to longer-distance-formed Cu-pairs and considered possible scenarios
depending on which Cu-pairs were taken into account. Also the analysis
of INS powder data led Zheludev {\it et al.}\cite{Zheludev_96}
 to conclude that this compound
should be described as a weakly interacting dimer system in spite
of the fact that the material shows a pronounced 1D arrangement
of magnetic ions. Which Cu ions form the pairs, however, 
couldn't be strictly defined from the analysis of the above experiments.

 It is the purpose of this letter to investigate in detail the
 singlet nature of this system and why it shows such a
behavior. In order to do that we have performed
 (i) {\it ab initio} calculations which give us information on the chemical
 bonding and the electronic structure and (ii) Quantum Monte Carlo computations
 which are used in order to calculate thermodynamical quantities like the
 susceptibility and magnetization in this material.

\begin{figure}[t]

\vspace*{0.1cm}

\hspace*{-0.5cm}\centerline{\hspace*{0cm}
\epsfig{file=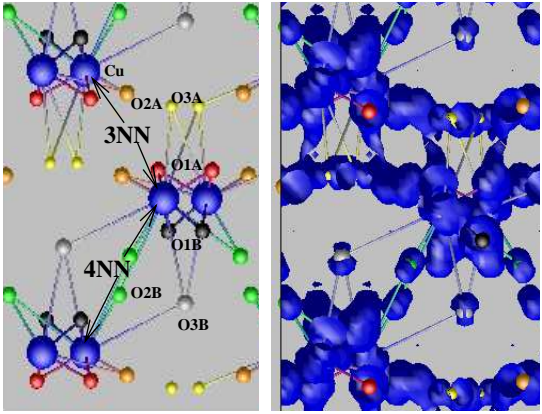,width=0.4\textwidth}
           }

\caption{\label{ED}
Electron density plot (right panel) 
for bands close to the Fermi level
in a plane perpendicular to the c-axis. Note the dominant electron 
density for the 3NN path with respect to the 4NN path. The left panel
 shows the
crystal structure in the corresponding plane. Big balls represent
Cu atoms while small balls represent O atoms.  The various non-equivalent
oxygens\protect\cite{Behruzi_86,explanation} are marked in the figure.
}
\end{figure}

{\bf Crystal structure}.- CaCuGe$_2$O$_6$ crystallizes in the 
monoclinic space-group P2$_1$/c with lattice parameters $a$ = 10.198 $\AA$,
$b$ = 9.209 $\AA$, $c$= 5.213 $\AA$, $\beta$ = 105.73$^o$, and it
contains four formula units per
primitive unit cell\cite{Behruzi_86}.
 The copper ions  are all equivalent and form Jahn-Teller distorted
octahedra with the surrounding oxygens and they build zig-zag chains
along the c-direction 
(see  Fig.\ \ref{structure1}(a)) by sharing an 
edge of the octahedra.  
Intercalated between these chains are the Ge and Ca ions.
The distances between two nearest neighbor (NN) Cu within the same 
chain is 3.072 $\AA$ and the Cu-O-Cu angles in these chains are
 92$^o$ for the Cu-O1A-Cu angle and 98$^{o}$ for the Cu-O1B-Cu angle
(see Fig.\ \ref{ED} and Ref.\ \cite{explanation} for the oxygen notation). 
The Cu-Cu distance within the 2nd NN pair which are also located in the
same zig-zag chain is 5.213 $\AA$. The 3rd and 4th NN are formed by Cu
atoms belonging to neighboring $bc$ planes (see Fig.\ \ref{structure1}).
 The distance between two 
neighboring $bc$ planes alternate between $d= 4.462 \AA$ and $d^{\ast} =
5.354 \AA$ giving rise to 3rd NN and 4th NN distances of 5.549 $\AA$
and 6.213 $\AA$ respectively.

\begin{figure}[t]

\vspace*{-0.5cm}

\centerline{\hspace*{0.5cm}
\epsfig{file=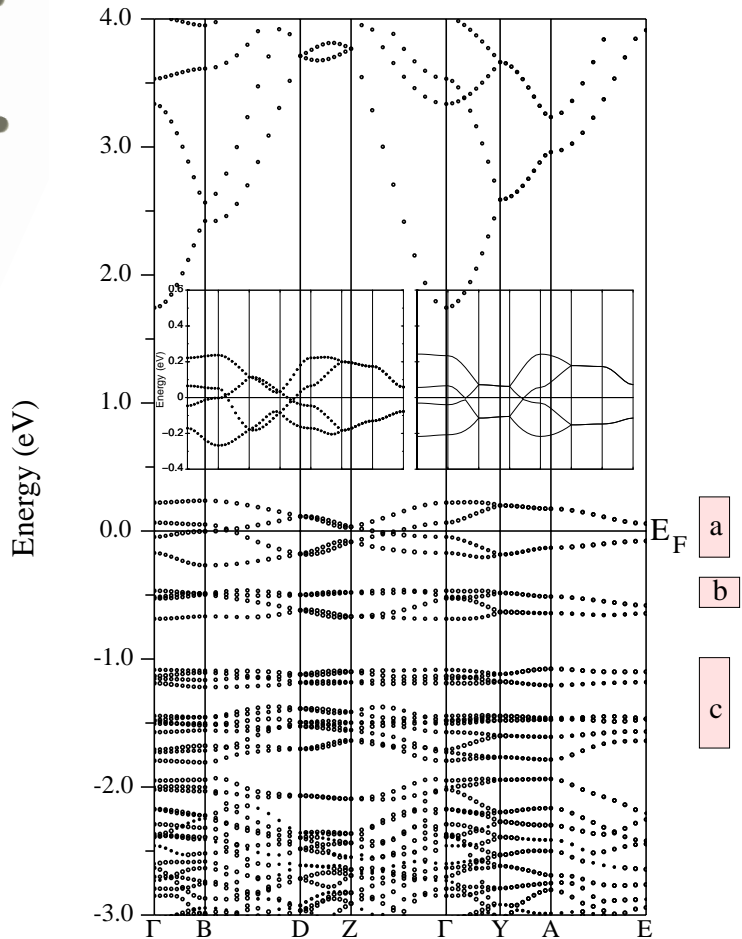,width=0.6\textwidth}
           }

\vspace*{-0.3cm}

\caption{\label{bands}DFT results for CaCuGe$_2$O$_6$. The path is 
along $\Gamma$=(0,0,0), B=(-$\pi$,0,0),
D=(-$\pi$,0,$\pi$),
Z=(0,0,$\pi$), $\Gamma$,
Y=(0,$\pi$,0), A=(-$\pi$,$\pi$,0)
E=(-$\pi$,$\pi$,$\pi$). Also shown in rectangles is the Cu band
character (in the
 local coordinate system) a= $d_{x^2-y^2}$, b=$d_{3z^2-1}$, c=$d_{xy}, d_{xz},
d_{yz}$ Shown in the inset are the four DFT-bands close to the
Fermi level (left panel) and the corresponding {\it downfolded}-TB
bands considering up to 4-th NN hopping (right panel) along the same path.
}
\end{figure}

{\bf Band-structure}.-
   We have performed an {\it ab initio} study
 based on the density-functional
 theory (DFT) in the generalized gradient approximation (GGA)\cite{Perdew_96}
 in order to derive  the electronic properties of CaCuGe$_2$O$_6$.
 Calculations have
  been performed  within the
 frame-work of both the full-potential
 linearized augmented plane wave (LAPW) method based on
 the WIEN97\cite{WIEN97} code and the
 linearized muffin tin orbital (LMTO)\cite{Andersen_75} method based
on the Stuttgart TBLMTO-47 code.
 The results of the band-structure calculations show
four narrow bands close to the Fermi level 
 with a bandwidth of $\sim$ 0.5 eV, having predominant Cu $d_{x^{2}-y^{2}}$
  character (in the local frame) admixed with oxygen contributions. 
 These bands are half-filled and strong
correlation effects should
explain the insulating groundstate in this compound.  These bands
 are separated by an energy gap of about 0.3 eV from the valence-band set
and a gap of about 2 eV from the conduction bands (see Fig.\ \ref{bands}).

{\bf Effective model}.- 
We have performed LMTO-based downfolding and tight-binding analysis
to derive the (single-particle)
tranfer integrals of an effective model for this system.

 In recent years a
new form of the LMTO method has been proposed and implemented
\cite{newlmto} which
allows to derive few-orbital effective Hamiltonians by
keeping only the relevant degrees of freedom and integrating out
the irrelevant ones. This procedure
 amounts to putting the inactive orbitals by downfolding
 in the tails of the
active orbitals kept in the basis.  As a result, this takes into
account the proper renormalization in the effective interactions
between the active orbitals. The effective interactions obtained
are therefore unique and given by the intervening interaction paths. 
Application of the downfolding procedure to CaCuGe$_{2}$O$_{6}$ 
to derive one orbital (namely $d_{x^{2}-y^{2}}$ per Cu atom) effective
model leads to the 
unambiguous result that the singlet pairs are formed by 3NN
Cu-pairs. The next important interactions are the 1NN providing 
a picture of two-dimensional interacting dimers in agreement
 with susceptibility data
and INS.  It is to be noted here that a simple-minded tight-binding
fitting method with one-band model is unable to distinguish the
relative importance between
3NN and 4NN pairs.
Either of the choice of dimer pairs would lead to similar band dispersions
though not to the same magnetic behavior.
 This dominant effect of the 3NN Cu-Cu interaction over
the 4NN Cu-Cu interaction is originated from  the specific orientation
of the intervening oxygen atoms between the Cu pairs and their
relative hybridization with Cu $d_{x^{2}-y^{2}}$ orbital guided by the
magnitude of Cu-O-Cu bond angles and the bond lengths. This is supported
by the 3D electron density plot shown in Fig.\ \ref{ED} near
the Fermi level in a plane perpendicular to the c-axis.
As it can be seen, the exchange paths
connecting two 3NN coppers are not equivalent to the exchange
paths connecting two 4NN coppers. For the considered 3NN Cu-Cu
pair the intervening oxygens (predominantly O2A and
O3A\cite{Behruzi_86,explanation}) provide a well-defined interaction path
 while
for the 4NN Cu-Cu pair the contribution of the intervening oxygens 
(predominantly O2B and O3B)
is much smaller leading to almost disconnected Cu$^{2+}$ ions.
This fact is also supported by the partial density of states (DOS) analysis
 at the Fermi level of the oxygens' contribution where we observe
that the contribution coming from O2B and O3B orbitals is almost
negligible  compared to that of O2A and O3A.

The {\it downfolded}-TB bands are represented by a Hamiltonian of
the form:
\begin{eqnarray}
H_{TB}=\sum_{(i,j)}t_{ij}(c^{\dagger}_i c_j + h.c.)
\end{eqnarray}
where the $t_{ij}$ are hopping matrix elements between Cu neighbors
 $i$ and $j$.
We have considered  
  hopping matrix elements up to fourth
 nearest neighbor (see Fig.\ \ref{structure1}) 
 $t_1$, $t_2$, $t_3$ and $t_4$ (the subindex denotes 1NN, 2NN, 3NN and 4NN 
 respectively). The  values derived for this effective model
 were
found to be  $t_1$=0.068eV, $t_2$=0.008eV, $t_3$=0.088eV and $t_4$=0.004eV.
  In the inset of Fig.\ (\ref{bands}) we show the comparison of the DFT-bands
 to the {\it downfolded}-TB bands considering $t_1$, $t_2$, $t_3$ and
$t_4$ hoppings only.  The band-splitting along the path
 $DZ$ is reproduced by considering longer-ranged hopping matrix elements.

 An estimate of the exchange integral related to the most dominating 3NN
interaction parameter $t_{3}$ can be obtained by using the relation 
$J_3 \sim 4 t_3^{2}/U$  where $U$ is the effective on-site Coulomb repulsion.
 A value of $U$ $\sim$ 4.2eV  was proposed for CuGeO$_3$
by mapping experimental data
onto one-band description\cite{Parmigiani_97}.  Assuming that this
 value is similar for CaCuGe$_2$O$_6$, we get a value of $J_3 \sim 86$K.

\begin{figure}[t]

\vspace*{0.1cm}

\centerline{\hspace*{0.0cm}
\epsfig{file=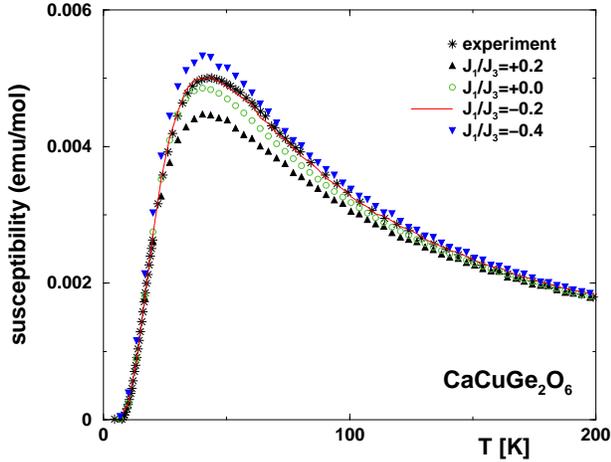,width=0.45\textwidth}
           }
\caption{\label{fig_chi}
Temperature dependence of the magnetic susceptibility
 for
CaCuGe$_2$O$_6$. Shown are the experimental\protect\cite{sas95} data (stars)
and the QMC results for the
model Eq.\ (\protect\ref{eq_model}) for  $J_3=67\,{\rm K}$ and
$J_1/J_3=0.2$ (up-triangles, $g=2.076$),
$J_1/J_3=0.0$ (open circles, $g=2.032$),
$J_1/J_3=-0.2$ (full line,   $g=2.032$) and
$J_1/J_3=-0.4$ (down-triangles,   $g=2.011$).
}
\end{figure}

{\bf Susceptibility and Magnetization}.- The description for
CaCuGe$_2$O$_6$ obtained from
the previous first principle calculation is that of a system of
 interacting dimers. In order to check this result we have analyzed
 both the susceptibility as well as the magnetization behavior of 
this material by considering the following model: 
\begin{equation}
H_{eff}\ =\ J_3 \sum_{<i,j>}S_i S_j +J_1 \sum_{(i,j)} S_i S_j~,
\label{eq_model}
\end{equation}
where $J_3$ and $J_1$ are the exchange integrals between 3NN and
1NN as illustrated in Fig.\ (\ref{structure1}).
The analysis of the model has been done by Quantum-Monte-Carlo (QMC)
simulations (stochastic series expansion~\cite{san99,dor01})
on $20\times40$ lattices. We found an optimal value for
$J_3=67\,{\rm K}$, very close to the value $68\,{\rm K}$ proposed
by Sasago {\it et al.}~\cite{sas95}. We find clear evidence
for a weak ferromagnetic inter-dimer coupling $J_1<0$, as
shown in Fig.\ (\ref{fig_chi}).
The optimal value is $J_1=-0.2 J_3$.
The $g$-factor was determined by fitting 
the QMC-results for the susceptibility,
$\chi^{(th)}=\langle \left(S^z-\langle S^z\rangle\right)^2\rangle$
with the experimental susceptibility $\chi$ (in [emu/mol])
at high temperatures via~\cite{joh00}
$\chi\equiv 0.375(g^2/J)\chi^{(th)}$.
The optimal $g$-factor was found to be $g=2.032$, close
to the value $2.07$ proposed earlier~\cite{sas95}.

\begin{figure}[t]

\vspace*{0.1cm}

\centerline{\hspace*{0.0cm}
\epsfig{file=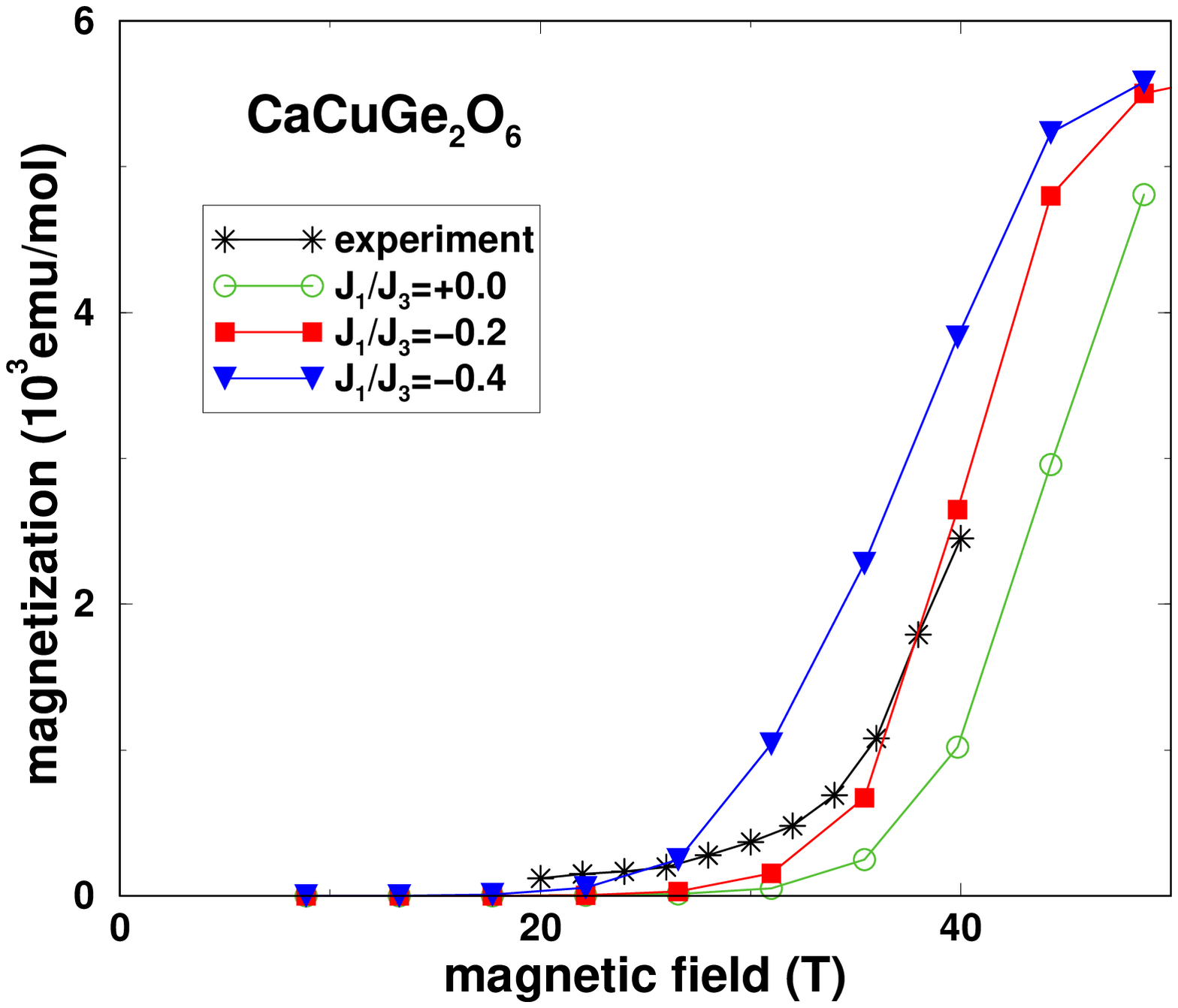,width=0.45\textwidth}
           }
\caption{\label{fig_mag}
 Magnetization for
CaCuGe$_2$O$_6$ at $T=4.2\,{\rm K}$. Shown are the
 experimental\protect\cite{sas95} data (stars)
and the QMC results for the
model Eq.\ (\protect\ref{eq_model}) for 
$J_1/J_3=0.0$ (open circles),
$J_1/J_3=-0.2$ (full squares) and
$J_1/J_3=-0.4$ (down-triangles) with the same
parameters as in Fig.\ (\protect\ref{fig_chi}).
}
\end{figure}

Sasago {\it et al.} measured the magnetization curve
for CaCuGe$_2$O$_6$ as a function of an applied magnetic field at
$T=4.2\,{\rm K}$.
With the stochastic series
expansion QMC-method  it is possible to simulate quantum-spin
models in an external field \cite{dor01}, although
these simulations become very hard near saturation.
We present the results of this simulation in
Fig.\ (\ref{fig_mag}), for the same parameters
we used for evaluating the susceptibility, together
with the experimental results. It is very reassuring
that $J_1=-0.2J_3$ is again the optimal parameter set.

The phase diagram of  model Eq.\ (\ref{eq_model}) is
very interesting by itself. In the limit
$J_3=0$ it consists of decoupled $J_1$-chains
(see Fig.\ (\ref{structure1}(a)))
either antiferro- or ferromagnetic for
$J_1>0$ or $J_1<0$ respectively. In both cases
the excitation spectrum is gapless, while it shows a
gap in the limit $J_1=0$  $J_3 \neq 0$ (dimer phase).
We therefore expect two quantum critical points
connecting these two regions. We have investigated
the phase-diagram and found the critical coupling
strength to be $J_1\simeq0.55 J_3$ and
$J_1\simeq-0.9 J_3$ respectively.  The parameter
range for CaCuGe$_2$O$_6$ is away from these two critical points.

{\bf Discussion and Conclusions}.-  the analysis of the electronic
 structure of CaCuGe$_2$O$_6$ by first principles calculations as
 well as the examination of susceptibility and magnetization data
by the QMC method leads to a unique description of this material
as a system of dimers formed by 3NN magnetic spin=1/2 Cu$^{2+}$ ions
with ferromagnetic (1NN) interdimer couplings.
The ferromagnetic nature of the 1NN coupling is more subtle to understand
than in other edge-sharing Cu-O systems\cite{Mizuno_98} since 
 due to the strong distorsion of the Cu
octahedron,
 this coupling proceeds through two
 pathes with angles 92$^o$ and 98$^o$. These two pathes 
 can provide in principle competing ferro-\cite{Good_55} and
 antiferromagnetic couplings
respectively. The analysis of the susceptibility and magnetization
 behavior shows nonetheless a clear
 effective ferromagnetic coupling.
%
 In this system, since the primary
 role is played by the  3NN Cu-pairs,
 the role of the possible
frustrating 2NN term to the 1NN interaction is secondary, unlike the case
of  the
related spin-Peierls material CuGeO$_3$.
 The strength of the magnetic exchange integrals has been obtained
 from the analysis of the susceptibility and the magnetization data
with very good agreement between these two measurements. DFT calculations
overestimate the value of $J_1$, which is not surprising\cite{livo},
 nonetheless the physical picture of this material is correctly
 provided by DFT+{\it downfolding}-TB analysis.

 In conclusion, by analizing the  electronic and
 magnetic properties of CaCuGe$_2$O$_6$ we have been able to
provide a physical picture of the system as well as a microscopic
description of the singlet
nature in this material. The combination of {\it ab initio} 
techniques with many-body
methods (QMC) can unambigously determine the microscopic behavior
of a system as has been shown for CaCuGe$_2$O$_6$ in particular. 

CaCuGe$_2$O$_6$ shows an interesting spin-gap
behavior and we believe that experiments under the application
of a magnetic field would be desirable to investigate the possible appearance
 of long-range order in the system. 
Moreover, the 2D interaction model provided by this system
 Eq.\ (\ref{eq_model})
 is interesting
 {\it per se} since it shows two quantum critical points as a function of
the strength quocient $J_1/J_3$.


%

\begin{thebibliography}{99}
\bibitem{explanation_1} The nearest neighbor coupling in CaV$_4$O$_9$
                      is argued to be small due to strong cancellation
                       among the various superexchange processes.
\end{thebibliography}
\end{document}